\documentclass[seceq]{ptptex}

\usepackage{graphicx}
\usepackage{wrapft}

\notypesetlogo                       %comment in if to eliminate PTPTeX 
\title{%        %You can use \\ for explicit line-break
Extended Pair Approximation of Evolutionary Game on Complex Networks%
}

\author{%       %Use \scshape  for the family name
Satoru \textsc{Morita}\footnote{
E-mail: morita@sys.eng.shizuoka.ac.jp}%
}

\inst{%     %Affiliation, neglected when [addenda] or [errata]
Department of Systems Engineering,  
Shizuoka University, \\ Hamamatsu 432-8561, Japan
}

\recdate{September 19, 2007}%            %Editorial Office will fill in this.

\abst{%         %this abstract is neglected when [addenda] or [errata]
We investigate how network structure influences 
evolutionary games on networks.
We extend the pair approximation to study 
the effects of degree fluctuation and clustering of the network. 
We find that a larger fluctuation of the degree is equivalent to
a larger mobility of the players.
In addition, a larger clustering coefficient is equivalent
to a smaller number of neighbors.}

\begin{document}

\maketitle

\section{Introduction}

Evolutionary games on networks have recently attracted attention 
in evolutionary biology, behavioral science and statistical 
physics.
In particular, an intensively studied question  is how 
network structure influences the
evolution of cooperative behavior in a dilemma situation
(for regular lattices\cite{nowak,nowak92,nakamaru,szabo,vukov,hauert}
and complex networks\cite{kim2002,ebel,masuda,tomochi,ohtsuki,zhong,tang,sagara}).
Nowak and May observed that cooperation behavior
can be  enhanced in the prisoner's dilemma game on a
lattice network.\cite{nowak92}
Contrastingly, Hauert and  Doebeli found that
network structure often inhibits cooperation behavior 
in the snowdrift game.\cite{hauert}
%Thus,
%The question of how to establish the theoretical formula
Because most of these papers are based on numerical simulations,
it is not clear how the network architecture
affects the evolution in general cases.
The purpose of this study is to establish a theoretical formula 
describing  the network effect in evolutionary game theory.

In this study, we focus on the influence 
of the average degree, degree fluctuations and 
clustering structure on the asymptotic result.
To analyze these effects, we apply the pair approximation technique.
The pair approximation is a useful tool for analysis of model ecosystems, 
because it can predict the population dynamics more accurately
than the mean-field approximation.\cite{matsuda,sato,morris,baalen}
The procedure for the pair approximation is often 
so complicated that it is carried out by numerical calculations.
Here, to obtain an analytical solution,
we propose novel procedures for the pair approximation.

%mobility 
%diffusion
%exchange
%static network 
\section{Games on Networks}
A game on network is defined as follows.
Let us consider a static network.
Each node of network is occupied by an individual.
Every individual plays games with its neighbors and
reproduces depending on the score of games.  
Furthermore, we introduce migration of 
the players on the static network to clear network effects.

In this study, we consider four types of networks.
First, we study the case of random regular graphs, in which all nodes have 
the same degree (the number of neighbors) $z$, 
and links are random (i.e. no correlations, clustering, etc.).\cite{bollobas}
We assume $z>2$ so that the networks are not divided into
many disconnected components.
Next, we expand the study to the case of networks in which
the degree is distributed:
Erd\H os-R\'enyi random graphs \cite{bollobas}
and Barab\'asi-Albert scale-free networks.\cite{barabasi}
Finally, to study the clustering effect, we consider
random regular graphs with a high level of clustering.\cite{kim}
%We consider game on a network with diffusion. 
%mobility of individuals 
%diffusion effect 

Consider a symmetric game with two strategies, $A$ and $B$,
with the payoff matrix
\[
\begin{array}{cc}
&
\begin{array}{cc}
A & B
\end{array}\\
\begin{array}{c}
 A\\
 B
\end{array}
&
\left(
\begin{array}{cc}
 a & b \\
 c & d
\end{array}
\right)
\end{array} .
\label{eq_gm}
\]
Here we assume $0<a,b,c,d<1$.
A player uses either strategy $A$ or $B$.
%The score of each player is calculated 
%by averaging the payoffs over all interactions
%with its neighbors.
The fitness of players $A$ and $B$ with 
$i$ neighbors with the same strategy ($A$ and $B$, respectively)
is given by 
%\begin{eqnarray}
% f_A(i)&=&1-w +w \left(a \frac{i}{z} +b \frac{z-i}{z}\right)\\
% f_B(i)&=&1-w +w \left(c \frac{z-i}{z}+d  \frac{i}{z}\right),
%\end{eqnarray}
\[
f_A(i)=1-w +w \left(a \frac{i}{z} +b \frac{z-i}{z}\right), \
f_B(i)=1-w +w \left(c \frac{z-i}{z}+d  \frac{i}{z}\right),
\]
where $w$ is a parameter that measures the intensity of selection.\cite{nowak}

Let us assume  the following update rule for evolutionary dynamics.
At each time step, with probability $1-q$,
reproduction process occurs as follows.
(i) An individual is selected
at random with a probability proportional to its fitness.
(ii) The selected individual is duplicated and it replaces a 
random neighbor.
With probability $q$,
a diffusion process occurs as follows.
(i) Two neighboring players are selected at random.
(ii) Their locations are exchanged.
The parameter $q$, which is between 0 and 1, measures
the intensity of the mobility of players.
In the limit  $q\rightarrow 1$, the dynamics becomes well mixed.
%The Moran process describes stochastic evolution of 
%a finite population of constant size.
%This process is called also birth-death process.

\section{Random regular graphs}

First, we present 
the theoretical results for random regular graphs
obtained with the pair approximation.
Let $X$ and $Y$ be the following conditional probabilities:
%that a random chosen neighbor of a player $A$ ($B$)
%has the same strategy $A$ ($B$).
%Thus we have 
\begin{equation}
X =  p_{{AA}}/p_{{A}},  \
Y =  p_{{BB}}/p_{{B}}.
\label{eq_p22}
\end{equation}
Here, $p_{*}$ is the concentration of player $*$, and 
$p_{**}$ represents the doublet density 
of two neighboring players. We have
$
p_{{A}}=p_{{AA}}+p_{{AB}}/2$ and 
$p_{{B}}=p_{{BB}}+p_{{AB}}/2 
$.
%From a simple algebra, we obtain the following set of equations:
%\begin{eqnarray}
%p_{AA} & = &  \frac{X(1-Y)}{2-X-Y} \\
%p_{AB} & = &  \frac{2(1-X)(1-Y)}{2-X-Y}\label{eq_p22} \\
%p_{BB} & = &  \frac{Y(1-X)}{2-X-Y}\\
%p_A & = &  \frac{1-Y}{2-X-Y} \label{eq_x} \\
%p_B & = &  \frac{1-X}{2-X-Y} \label{eq_y} .
%\end{eqnarray} 
%Here $p_A$ and $p_B$ represent the densities of $A$ and $B$ players
%, respectively .
In the pair approximation, 
the system can be described by the two variables $X$ and $Y$ alone:
\begin{eqnarray}
p_A & = &  \frac{1-Y}{2-X-Y} ,\nonumber \\
p_B & = &  \frac{1-X}{2-X-Y} ,\nonumber \\
p_{AA} & = &  \frac{X(1-Y)}{2-X-Y} ,
\label{eq_x}\\
p_{AB} & = &  \frac{2(1-X)(1-Y)}{2-X-Y} ,\nonumber \\
p_{BB} & = &  \frac{Y(1-X)}{2-X-Y}\nonumber.
\end{eqnarray}
The probability that a player $A$ has 
$i$ neighbors with strategy $A$ is given by
\[
p_{{A}}(i)=
 \left(
 \begin{array}{c}
  z\\i
 \end{array}
 \right)
 X^i (1-X)^{z-i} .
\] 
%The probability that a player with strategy B has 
%$i$ neighbors with strategy B is given by 
%\[
%p_B(i)=
% \left(
% \begin{array}{c}
%  z\\i
% \end{array}
% \right)
% Y^i (1-Y)^{z-i} ,
%\]
The strategy $A$ replaces the strategy $B$ only if
the player selected to reproduce is $A$ and the replaced neighbor is $B$.
The probability that  this event occurs is given by
\begin{equation}
\begin{array}{lcl}
 P_{{B}\rightarrow {A}}&=&\displaystyle
\frac{p_{{A}}}{\Phi} \sum_{i=0}^{z}\frac{z-i}{z}
f_{{A}}(i)p_{{A}}(i)
\\ 
&=& \displaystyle \frac{p_{{AB}}}{2\Phi}  \left\{
1-w+w [b+(1-1/z)(a-b)X]
\right\},
\end{array}
\label{eq_pba}
\end{equation}
where we have used Eqs.~(\ref{eq_p22}) and (\ref{eq_x}) 
to derive the final part.
Here $\Phi$ is a normalization constant,
which is given by the average fitness over all individuals:
\begin{equation}
\Phi=1-w+w\{p_A [b+(a-b) X]+p_B[c+(d-c)Y]\} . 
\label{eq_phi}
\end{equation}
In the same way, 
the probability that the strategy B replaces the strategy $A$
is given by
\begin{equation}
\begin{array}{lcl}
 P_{{A}\rightarrow {B}}
%&=&\displaystyle
%\frac{p_{{B}}}{\Phi}  \sum_{i=0}^{z}\frac{z-i}{z}
%f_{{B}}(i)p_{{B}}(i)
%\\ 
&=&\displaystyle
\frac{p_{{AB}}}{2\Phi}  \left\{
1-w+w [c+(1-1/z)(d-c)Y]
\right\}. 
\end{array} 
\label{eq_pab}
\end{equation}
From Eqs.~(\ref{eq_pba}) and (\ref{eq_pab}),
we obtain
\begin{equation}
\dot{p}_A=\frac{p_{{AB}}}{2\Phi} w [b-c +(1-1/z)(a-b)X-(1-1/z)(d-c)Y] .
\label{eq_pp1}
\end{equation}
The necessary condition for equilibrium 
is $\dot{p}_A=0$.
This  condition is simplified as
\begin{equation}
(a-b)X+(c-d)Y=\frac{c-b}{1-1/z}
\label{eq_eq1}
\end{equation}
From Eqs.~(\ref{eq_pba}), (\ref{eq_phi}), (\ref{eq_pab}) and (\ref{eq_eq1}),
we obtain  the following approximate relation:
\begin{equation}
P_{{B}\rightarrow {A}}=P_{{A}\rightarrow {B}}
= p_{{AB}}/2 + O(w/z).
\label{eq_appro}
\end{equation}
\begin{figure}[t]
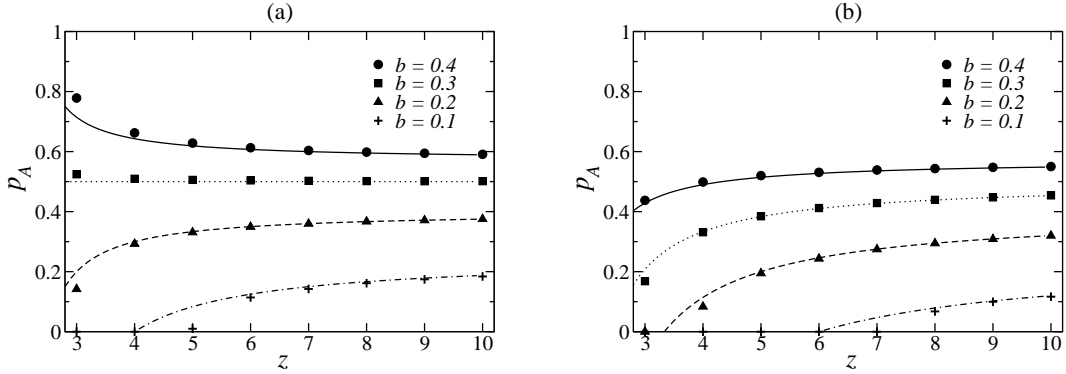

\includegraphics[width=65mm,clip]{figure1a} \hfill%
\includegraphics[width=65mm,clip]{figure1b}%
\caption{
The density $p_{{A}}$ of player $A$  plotted 
as a function of the degree $z$ 
for a random regular graph 
(a) $q=0$, (b) $q=0.2$.
The game parameters are set as 
$b=0.1$, $0.2$, $0.3$ and $0.4$ from bottom to top
for fixed $a=0.7$, $c=1$, $d=0$ and $w=0.5$.
The total number of players is fixed to 10000. 
In all simulations, $p_A$ is obtained by averaging over the last 10,000 
time steps after the first 10,000 ones, 
and each data point results from 10 different network realizations.
The curves represent
the theoretical predictions (\ref{eq_main}).
\label{fig_1}}
\end{figure}
In addition, through the reproduction process,
the rate of change of the doublet density is given by
\begin{equation}
\begin{array}{lcl}
P_{{AB}\rightarrow {AA}} & = 
& [1+(z-1)(1-Y)]P_{{B}\rightarrow {A}}\\
%P_{BB\rightarrow AB} & = & (z-1)Y P_{B\rightarrow A}\\
%P_{AB\rightarrow BB} & = & [1+(z-1)(1-X)]P_{A\rightarrow B}\\
P_{{AA}\rightarrow {AB}} & =
 & (z-1)XP_{{A}\rightarrow {B}} .
\end{array}
\label{eq_p1}
\end{equation}
Here, we have used the fact that the replaced player
%, which changes the strategy, 
has $z-1$ neighbors, other than  the player selected
to reproduce.
Then, through the diffusion process,
the rate of change of the doublet density is given by
\begin{eqnarray}
P'_{{AB}\rightarrow {AA}} & = &  (z-1)(1-Y)p_{{AB}}, \nonumber \\
%P'_{BB\rightarrow AB} & = &  Y(z-1)p_{AB}\\
%P'_{AB\rightarrow BB} & = &  (1-X)(z-1)p_{AB}\\
P'_{{AA}\rightarrow {AB}} & = &  (z-1)X p_{{AB}} .
\label{eq_p2}
\end{eqnarray}
Here, we have used the fact that
each node of the doublet has  $z-1$ links  
excluding the link between the doublet.
We have
\begin{equation}
\dot{p}_{AA}=
(1-q)P_{{AB}\rightarrow {AA}}+
q P'_{{AB}\rightarrow {AA}}-
(1-q)P_{{AA}\rightarrow {AB}}-
q P'_{{AA}\rightarrow {AB}}.
\label{eq_pp2}
\end{equation}
The equilibrium state  must satisfy $\dot{p}_{AA}=0$.
%\begin{eqnarray}
%(1-q)P_{AB\rightarrow AA}+q P'_{AB\rightarrow AA}\nonumber \\
%=(1-q)P_{AA\rightarrow AB}+qP'_{AA\rightarrow AB}\nonumber\\
%(1-q)P_{AB\rightarrow BB}+q P'_{AB\rightarrow BB}\nonumber \\
%=(1-q)P_{BB\rightarrow AB}+qP'_{BB\rightarrow AB} \label{eq_pp2}
%\end{eqnarray}
%\begin{equation}
%(1-q)P_{{AB}\rightarrow {AA}}+
%q P'_{{AB}\rightarrow {AA}}=
%(1-q)P_{{AA}\rightarrow {AB}}+
%q P'_{{AA}\rightarrow {AB}} 
%\end{equation}
Using Eqs.~(\ref{eq_appro}), (\ref{eq_p1}) and (\ref{eq_p2}),
we obtain 
\begin{equation}
X+Y=1+\frac{1-q}{(1+q)(z-1)}
\label{eq_XY}
\end{equation}
This result is an approximation valid  for $qw/z\ll 1$  .
The equilibrium values of $X$ and $Y$ are obtained
by solving eqs.~(\ref{eq_eq1}) and (\ref{eq_XY}).
From Eq.~(\ref{eq_x}),
the equilibrium density of player $A$ is given by 
\begin{equation} 
%{p_{{A}}}_*=\frac{(1+q)(b-d)z-(c-a+b-d)-q(c+a-b-d)}
%{(c-a+b-d)[(1+q)z-2]} .
{p_{A}}_*= \frac{b-d}{c-a+b-d}
-\frac{c-a-b+d+q(c+a-b-d)}{(c-a+b-d)[(1+q)z-2]}
\label{eq_main}
\end{equation}
The stability is determined by calculating the
Jacobi matrix of the dynamics of $X$ and $Y$ from (\ref{eq_pp1}) and 
(\ref{eq_pp2}).
After using (\ref{eq_appro}) and some algebra, we obtain that 
if $c-a+b-d>0$ and $0<{p_{A}}_*<1$  then ${p_{A}}_*$
is stable. 
The results are summarized in Table I.
\begin{table}[t]
\caption{Equilibrium and its stability.
The solid and dotted curves indicate stable and unstable equilibrium, 
respectively. Here, we have
$z_0=\frac{c-a+b-d+q(c+a-b-d)}{(b-d)(1+q)}$
 and $z_1=\frac{c-a+b-d-q(c+a-b-d)}{(c-a)(1+q)}$.
}
\begin{center}
\begin{tabular}{|c|c|c|c|l|}
\hline
$c-a$ & $b-d$ & $c-a+b-d$ & 
$c-a-b+d$
 & equilibrium\\
& & & $+q(c+a-b-d)$ &\\
\hline
\hline
$+$ & $+$ & $+$ & $+$ & 
%(\ref{eq_main}) is stable  for $z>z_0$,\\
%& & & &  
\includegraphics[width=16mm,clip]{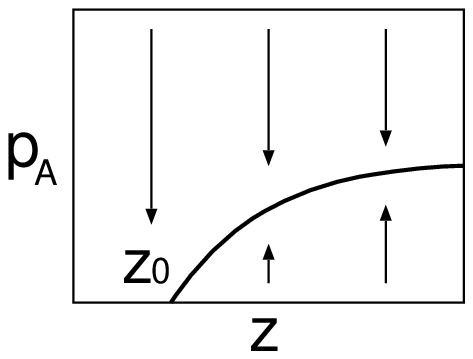}
%$p_A=0$ is stable for $z<z_0$.
\\
\hline
$+$ & $+$ & $+$ & $-$ &  
%(\ref{eq_main}) is stable for $z>z_1$,\\
%& & & &  
\includegraphics[width=16mm,clip]{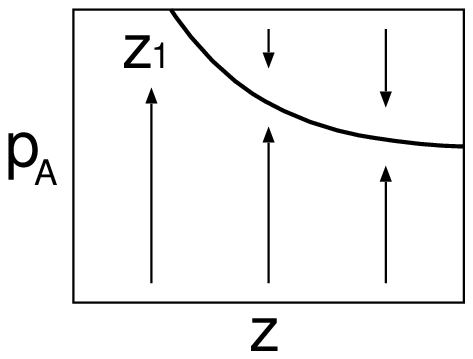}
%$p_A=1$ is stable for $z<z_1$.
\\
\hline
$-$ & $-$ & $-$ & $+$ & 
%(\ref{eq_main}) is unstable for $z>z_1$,\\
\includegraphics[width=16mm,clip]{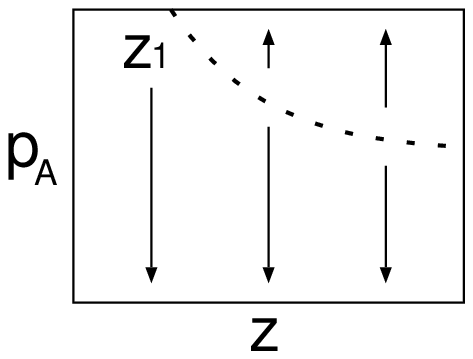}\\
%& & & &  $p_A=1$ is stable for $z>z_1$,\\
%& & & &  $p_A=0$ is always stable.\\
\hline
$-$ & $-$ & $-$ & $-$ & %
%(\ref{eq_main}) is unstable  for $z>z_0$, \\
\includegraphics[width=16mm,clip]{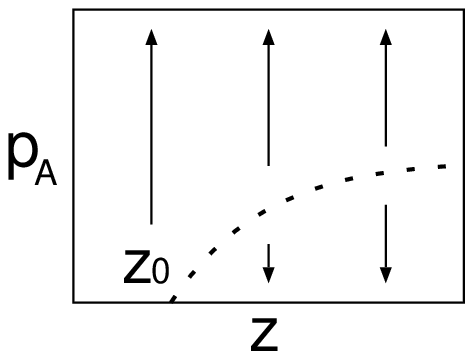}\\
%& & & &  $p_A=0$ is stable for $z>z_0$,\\
%& & & &  $p_A=1$ is always stable.\\
\hline
$+$ & $-$ & $+$ & $+$ &  
\includegraphics[width=16mm,clip]{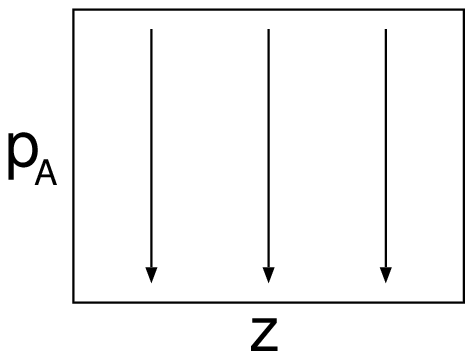}\\
%(\ref{eq_main}) is always smaller than 0,\\
%& & & & $p_A=0$ is always stable.\\
\hline 
$+$ & $-$ & $+$ & $-$ & 
%(\ref{eq_main}) is stable  for $z_1<z<z_0$,\\
\includegraphics[width=16mm,clip]{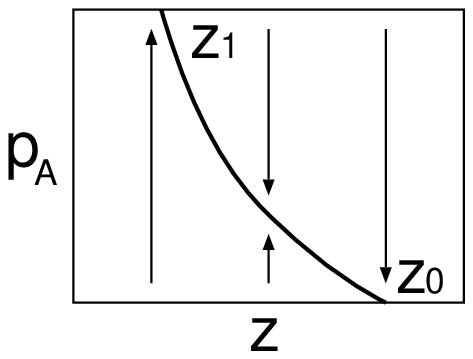}\\
%& & & &  $p_A=0$ is stable for $z>z_0$,\\
%& & & &  $p_A=1$ is stable for $z<z_1$.\\
\hline
$-$ & $+$ & $+$ & $+$ & 
%(\ref{eq_main}) is stable  for $z_0<z<z_1$, \\
\includegraphics[width=16mm,clip]{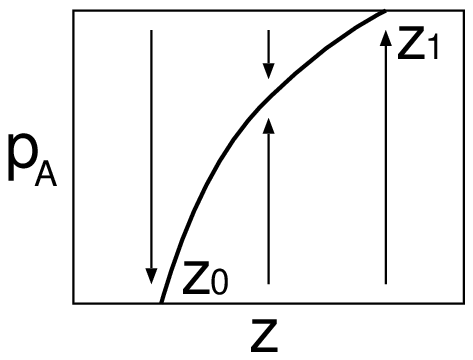}\\
%& & & &  $p_A=0$ is stable for $z<z_0$,\\
%& & & &  $p_A=1$ is stable for $z>z_1$.\\
\hline
$-$ & $+$ & $+$ & $-$ & 
%(\ref{eq_main}) is always larger than 1,\\
\includegraphics[width=16mm,clip]{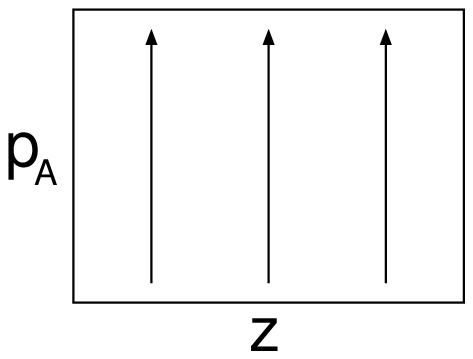}\\
%& & & & $p_A=1$ is always stable.\\
\hline
$+$ & $-$ & $-$ & $+$ & 
% (\ref{eq_main}) is always larger than 1,\\
\includegraphics[width=16mm,clip]{pmpp}\\
%& & & & $p_A=0$ is always stable.\\
\hline 
$+$ & $-$ & $-$ & $-$ & 
%(\ref{eq_main}) is unstable  for $z_0<z<z_1$,\\
\includegraphics[width=16mm,clip]{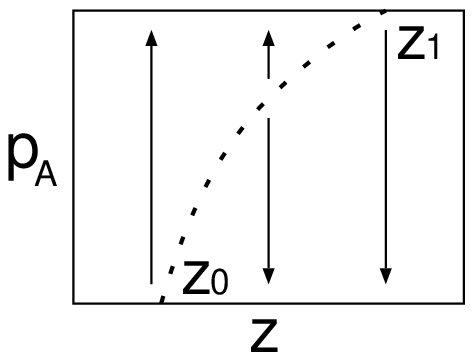}\\
%& & & &  $p_A=0$ is stable for $z>z_0$,\\
%& & & &  $p_A=1$ is stable for $z<z_1$.\\
\hline
$-$ & $+$ & $-$ & $+$ &
% (\ref{eq_main}) is unstable  for $z_1<z<z_0$, \\
\includegraphics[width=16mm,clip]{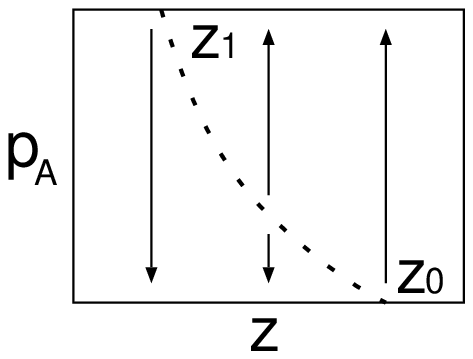}\\
%& & & &  $p_A=0$ is stable for $z<z_0$,\\
%& & & &  $p_A=1$ is stable for $z>z_1$.\\
\hline
$-$ & $+$ & $-$ & $-$ &  
%(\ref{eq_main}) is always smaller than 0,\\
\includegraphics[width=16mm,clip]{mppm}\\
%& & & & $p_A=1$ is always stable.\\
\hline
\end{tabular}
\end{center}
\end{table}

For example, when $c>a$ and $b>d$ (the chicken game),
the equilibrium ${p_A}_*$ is stable for large $z$ 
(see the first and second rows in Table I). 
When $c<a$ and $b<d$ (the assurance game),
the equilibrium  ${p_A}_*$ is unstable and thus
the density of players should approach to 0 or 1, depending on the
initial conditions (see the third and fourth rows in Table I).
%When $a>c$ and $b>d$ ($a<c$ and $b<d$),
Furthermore, when  $b<d<a<c$ (the prisoner's dilemma game),
either ${p_{{A}}}_*<0$ (see the fifth row in Table I) or 
${p_{{A}}}_*>1$ (see the ninth row in Table I).
Thus, all players should use strategy B
(i.e. uncooperative behavior) ultimately
and strategy $A$ (i.e. cooperative behavior) is never sustainable.
In the limit $z\rightarrow \infty$, Eq.~(\ref{eq_main}) approaches 
$(b-d)/(c-a+b-d)$, which is the Nash equilibrium in 
conventional game theory.
%Without diffusion $q=0$, we have
%\begin{equation}
%{p_{{A}}}_* = \frac{b-d}{c-a+b-d}+\frac{1}{z-2}\ \frac{a-c+b-d}{c-a+b-d} .
%\end{equation}
%On the other hand, 
In the limit $q\rightarrow 1$,
Eq.~(\ref{eq_main}) corresponds to the mean-field approximation.
%\begin{equation}
%{p_{{A}}}_* = \frac{b-d}{c-a+b-d}+\frac{1}{z-1}\ \frac{b-c}{c-a+b-d} .
%\label{eq_mf}
%\end{equation}
%In fact, substituting $q=1$ into (\ref{eq_XY}) yields $X+Y=1$,
%and substituting $X+Y=1$ into (\ref{eq_p22}) and (\ref{eq_x}) 
%yields $p_{{AB}}=p_{{A}} p_{{B}}$.
Note that the mean-field approximation does not 
coincide with the Nash equilibrium, because
the number of opponent players is restricted. 
Figure 1 shows that this approximation
agrees very well with the numerical simulations. 

\section{Degree fluctuation}
Let us now consider the case 
in which the degree is 
not uniform but exhibits a distribution $\rho(k)$.
Then, the average degree is written 
\begin{equation}
z=\langle k \rangle =\sum_k k \rho(k) .
\end{equation} 
In this case, we need to extend the pair approximation.
Here, we do not take degree correlation into account.
Furthermore, we assume that the density $p_A$ and 
the conditional probabilities $X$ and $Y$
do not depend on the degree.
Without this assumption, the relation (\ref{eq_x}) no longer hold, and thus
we need more than two variables to describe the system.
Although three variables (e.g. $p_A$, $p_{AA}$ and $p_{AB}$) are often used 
to account for the dependence of $p_A$ on the degree\cite{morris,baalen},
the outcome is too complicated to give clear insight.
In addition, the approximation with three variables 
yields results that  are  quantitatively similar 
to the approximation with two variables, because 
the dependence of $p_A$ on the degree is weak, as seen below.
Accordingly, we adopt the approximation with two variables $X$ and $Y$.

If a link is selected at random,
the distribution of 
the degree of the nodes to which the particular link leads is
not $\rho(k)$ but rather $k\rho(k)$.\cite{molloy,newman}
Thus, the average degree of the player replaced in the
reproduction process and 
the two players of the exchanged doublet in the diffusion process is given by
\begin{equation}
\frac{ \sum_k k^2 \rho(k)}{\sum_k k \rho(k)}=
\frac{\langle k^2 \rangle}{\langle k \rangle} .
\label{eq_k2k}
\end{equation}
In this case,
we should use (\ref{eq_k2k}) instead of $z$ in (\ref{eq_p1}) and
(\ref{eq_p2}) to calculate the equilibrium.
As a result, instead of (\ref{eq_XY}), we have 
\begin{equation}
X+Y=1+\frac{1-q}{(1+q)(\langle k^2 \rangle/\langle k\rangle-1)} .
\label{eq_XY2} 
\end{equation}
Equation (\ref{eq_XY2}) can be obtained by substituting  
the effective mobility 
\begin{equation}
q'= q+ \frac{1-q^2}{\kappa+q} 
\label{eq_qq}
\end{equation}
for $q$ in (\ref{eq_XY}),
where $\kappa=(\langle k^2 \rangle
+\langle k \rangle^2-2\langle k \rangle)/
(\langle k^2 \rangle-\langle k \rangle^2) $.
%\begin{equation}\kappa=\frac{\langle k^2 \rangle
%+\langle k \rangle^2-2\langle k \rangle}
%{\langle k^2 \rangle-\langle k \rangle^2} > 1.
%\label{eq_ka}
%\end{equation}
The parameter $\kappa$ is larger than 1 for $z=\langle k\rangle>2$, 
and thus it decreases with the variance of the degree distribution.
It is obvious that the additional part, $(1-q^2)/(\kappa+q)$,
in (\ref{eq_qq})
is  positive (because $0<q<1$) and a decreasing function of $\kappa$.
Consequently, increasing the variance of the degree
is equivalent to increasing the mobility, $q$.
\begin{figure}[t]
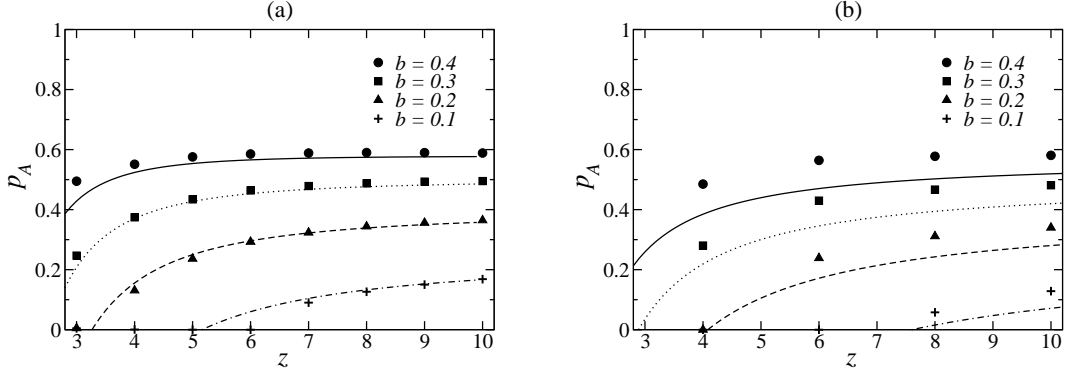

\includegraphics[width=65mm,clip]{figure2a} \hfill%
\includegraphics[width=65mm,clip]{figure2b}%
\caption{
The density of player $A$ is plotted 
as a function of the average degree $z$ 
for Erd\H os-R\'enyi random graphs (a)
and for Barab\'asi-Albert scale-free networks (b).
The curves represent the theoretical prediction.
Here $q=0$, and the other parameters are the same as Fig.~1.
\label{fig_2}}
\end{figure}

For examples, 
we consider Erd\H os-R\'enyi random graphs
and Barab\'asi-Albert scale-free networks.
For Erd\H os-R\'enyi random graphs,
the degree follows a Poisson distribution,
$\rho(k)=e^{-z}z^k/k!$,
which leads to $\langle k^2 \rangle=z(z+1)$.
Thus, we obtain $\kappa=2z-1$.
For Barab\'asi-Albert scale-free networks,
the degree follows a power law distribution, $\rho(k)\propto k^{-\gamma}$,
with the exponent $\gamma=3$.
In this case, $\langle k^2 \rangle \simeq z^2 \log N /4$
is obtained at leading order.\cite{pastor-satorras}
Thus, we obtain $\kappa=(z\log N+4z-8)/(z\log N-4z)$.
In the limit $N\rightarrow \infty$, we have $q'\to 1$,
which means that  the result approaches that of 
the mean-field approximation.
For Erd\H os-R\'enyi random graphs, the theory
agrees well with the numerical simulations [see Fig.~2(a)].
For Barab\'asi-Albert scale-free networks, however,
the agreement is not so good [see Fig.~2(b)].
This deviation mainly results from the fact that
we ignored dependence of $X$ and $Y$ on the degree in this approximation.
Figure 3 shows the numerical result,
where $p_A$ appears to be independent of the degree $k$,
but $X$ and $Y$ decrease with $k$.
Thus, a node with smaller degree tends to 
have homogeneous neighbors.

\begin{figure}[t]
\begin{minipage}{0.48\hsize}
\includegraphics[width=63mm,clip]{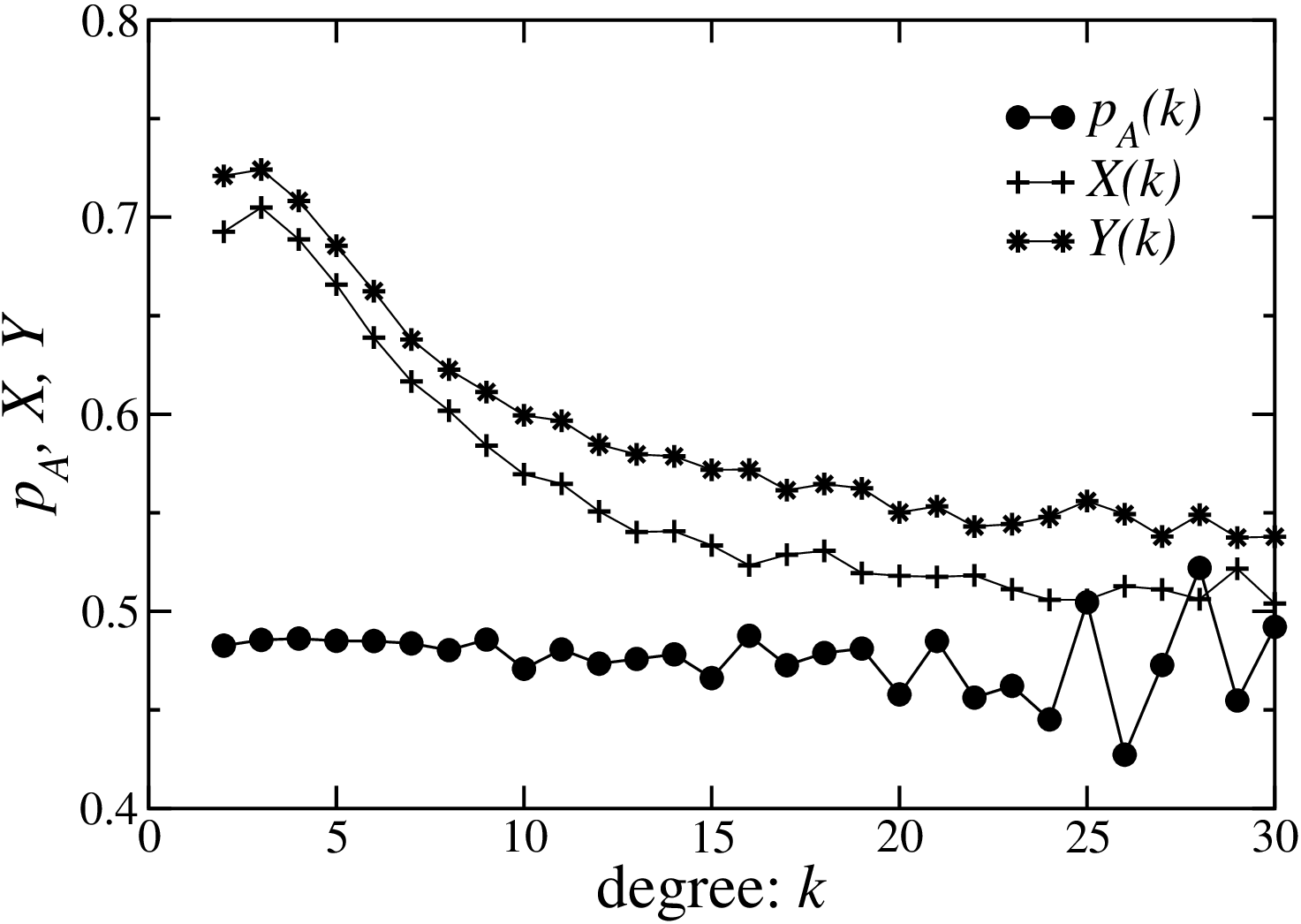}
\caption{
The density $p_{A}$ of player $A$ 
and the conditional probabilities $X$ and $Y$
are plotted as  functions of the degree $k$
for Barab\'asi-Albert scale-free networks.
Here $a=0.7$, $b=0.4$, $c=1$, $d=0$, $w=0.5$, $q=0$ and $z=4$.
The curves are a guide for the eyes.
\label{fig_3new}}
\end{minipage}
\hspace{0.03\hsize}
\begin{minipage}{0.48\hsize}
\includegraphics[width=63mm,clip]{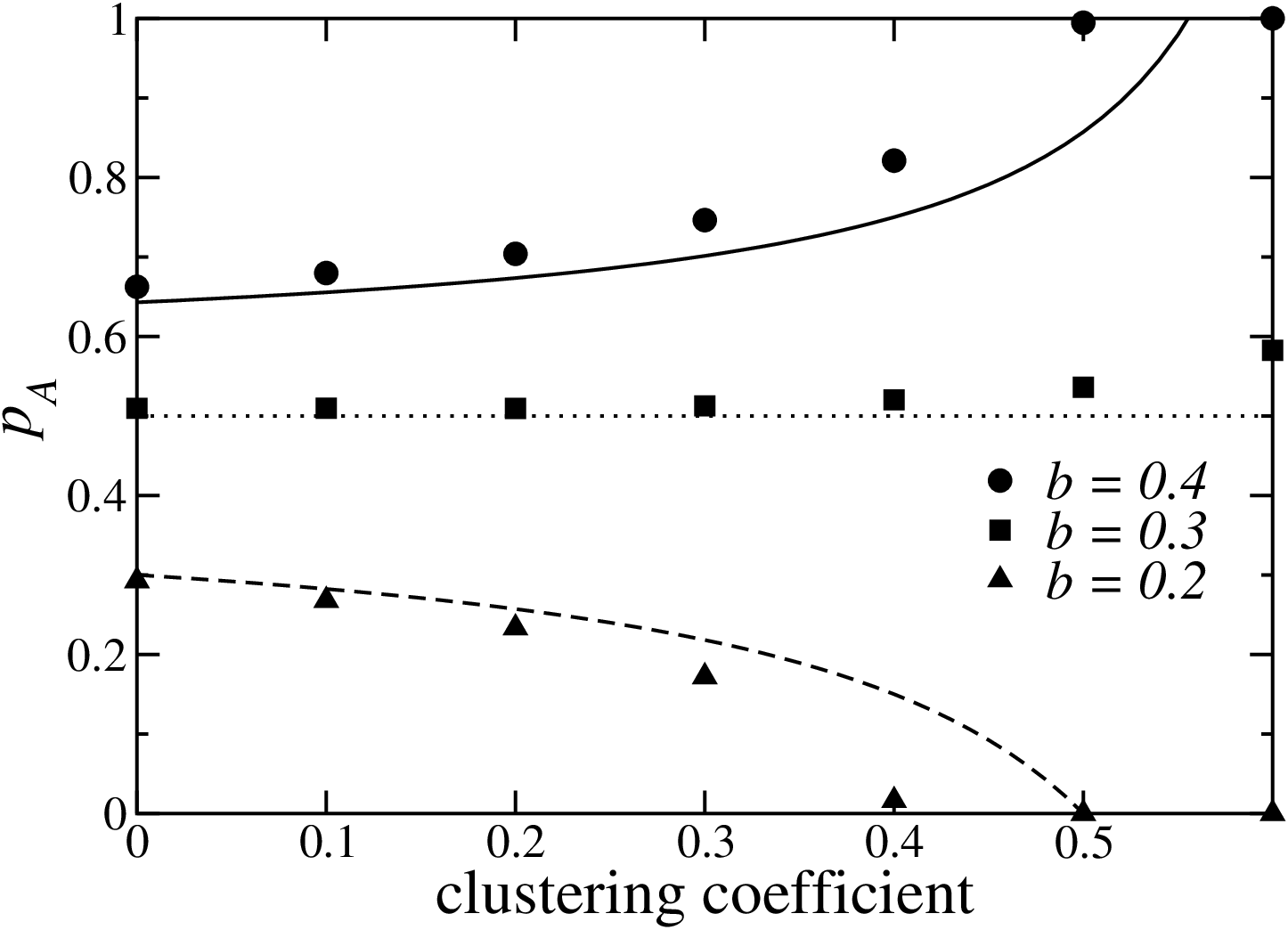}
\caption{
The density of player $A$ is plotted 
as a function of the clustering coefficient $C$
for networks whose nodes have fixed degree $z=4$.
Here $q=0$ and the other parameters are the same as Fig.~1.
\label{fig_3}}
\end{minipage}
\end{figure}

\section{Clustering effect}
We now turn to a study of the clustering effect.
For simplicity, we return to the case in which the degree is uniform. 
To this point, we have used network with 
a very small clustering coefficient.
%Let us study networks with higher clustering coefficient.
The clustering coefficient quantifies
the probability that two vertices that are connected to
the same node are also connected.\cite{watts}
Thus, a network with a large clustering coefficient has 
many triangles, i.e., loop-like triplets.
Ingnoring the triplet correlation without pair correlation biases, 
we assume that the density of three players on a triangle follows
Kirkwood superposition approximation:\cite{kirkwood,matsuda2000,keeling}
\[
 p_{{AAB}}:
 p_{{ABB}} \sim \frac{p_{{AA}}p_{{AB}}^2}
{p_{{A}}^2 p_{{B}}}:
 \frac{p_{{AB}}^2 p_{{BB}}}{p_{{A}}p_{{B}}^2}=  X:Y .
\]
Note that in the original Kirkwood superposituon approximation, 
the normalization condition is violated.

Recall that 
the configuration of strategies changes only if 
two neighboring players have different strategies (i.e. $A$ and $B$).
The probability that these two neighboring players 
and another neighbor of one of them
compose a triangle is given by $C$.
In this case, the probabilities that the third player
is $A$ and $B$ are $p_{{AAB}}$
and $p_{{ABB}}$, respectively.
Accordingly, instead of  
$X$ and $Y$ in Eqs.~(\ref{eq_eq1}) and (\ref{eq_XY}),
we should use 
\begin{equation}
 \displaystyle X'= X (1-C)+ C \frac{X}{X+Y}, \
 \displaystyle Y'= Y (1-C)+ C \frac{Y}{X+Y} .
\label{eq_XXYY}
\end{equation}
Here we should note that a pair of configurations 
$p_{AAB}$ and $p_{ABB}$ are normalized.
In other studies, 
all configurations $p_{AAA}$, $p_{AAB}$, $p_{ABB}$ 
and $p_{BBB}$ have been normalized  for similar approximations.\cite{baalen,keeling}
Our approximation is better.

The formulation obtained using the replacement (\ref{eq_XXYY}) can be also
obtained by substituting  
$z'$ and  $q'$ for $z$ and $q$ in (\ref{eq_eq1}) and (\ref{eq_XY})
as follows:
\[
 \begin{array}{ccl}
 q'& =& 
 \displaystyle q\left\{1- \frac{(1-q^2)(z-1)}{(z-2)q+z}C+O(C^2)\right\}, \\
 z'& =&
 \displaystyle z\left\{1- \frac{(1-q)(z-1)}{(z-2)q+z}C+O(C^2)\right\} .
 \end{array} 
\]
Thus, an increase of the clustering coefficient
is equivalent to a decrease of $z$ and $q$.
In particular, when $q=0$,
this substitution is simplified exactly as 
\begin{equation}
z'=z-C(z-1)
\end{equation}
In this case, the clustering effect is 
equivalent to the effect of decreasing the number $z$ 
of neighbors.
We present numerical results in Fig.~4.
Here, to introduce the clustering structure into the random regular graphs,
we used the edge exchange method.\cite{kim}
In this method, two links are selected randomly, and 
they are rewired only when the new network configuration 
is connected and has a larger clustering coefficient.
Figure 4 shows that
our approximation agrees well with the numerical simulations. 
The deviation seen in the region of large $C$
may be due to the fact that we ignored the effect from loops with
more than three nodes.

%\begin{wrapfigure}{l}{65mm}
%\centerline{\includegraphics[width=63mm,clip]{figure3new}}
%\caption{
%The density $p_{A}$ of player $A$ 
%and the conditional probability $X$ and $Y$
%are plotted as a function of the degree $k$
%Here $a=0.7$, $b=0.4$, $c=1$, $d=0$, $w=0.5$, $q=0$ and $z=4$.
%\label{fig_3new}}
%\end{wrapfigure}
%\begin{wrapfigure}{r}{65mm}
%\centerline{\includegraphics[width=63mm,clip]{figure3}}
%\caption{
%The density of player $A$ is plotted 
%as a function of the clustering coefficient $C$
%for networks whose nodes has fixed degree $z=4$.
%The other parameters are same to Fig.~1.
%\label{fig_3}}
%\end{wrapfigure}
 
\section{Summary}
In conclusion, we have studied the network effect 
in general $2\times2$ game by using the pair approximation.
First, for random regular graphs, our theoretical results
are presented in (\ref{eq_main}) and Table I.
Then, by the extended pair approximation, 
we developed a theory
for networks with degree fluctuation
and  networks with large clustering.
It was found that a fluctuation of the degree has the same effect as 
an increase in the mobility $q$, and
a clustering structure has the same effect 
as a decrease in the number $z$ of neighbors.
Real social networks are more complex than the networks
used in our numerical simulations,
because they have 
degree correlation\cite{maslov}, 
hierarchy and community structures.\cite{watts2}
We believe our method will be useful as a first step in analyzing 
such complicated situations. 
Furthermore, many other update rules for evolutionary dynamics 
are proposed.\cite{hauert,nakamaru}
Some behavior seen in this paper depend on the update rule.
The investigation of the dependence on 
the update rule is a future project.

\section*{Acknowledgements}
%We would like to thank ...........
This research was supported by the Japan Society of Promotion of
Science under the contract number CE19740234.
Some of the numerical calculations were carried out on 
machines at YITP of Kyoto University.

%\appendix
%\section{First Appendix} %Empty argument \section{} yields `Appendix'. 
%
%\section{Second Appendix}

\end{document}